\documentclass{article}

\usepackage{authblk}
\usepackage{graphicx}
\usepackage{listings, xcolor}
\usepackage{setspace}

\definecolor{verylightgray}{rgb}{.97,.97,.97}

\lstdefinelanguage{Solidity}{
	keywords=[1]{anonymous, assembly, assert, balance, break, call, callcode, case, catch, class, constant, continue, constructor, contract, debugger, default, delegatecall, delete, do, else, emit, event, experimental, export, external, false, finally, for, function, gas, if, implements, import, in, indexed, instanceof, interface, internal, is, length, library, log0, log1, log2, log3, log4, memory, modifier, new, payable, pragma, private, protected, public, pure, push, require, return, returns, revert, selfdestruct, send, solidity, storage, struct, suicide, super, switch, then, this, throw, transfer, true, try, typeof, using, value, view, while, with, addmod, ecrecover, keccak256, mulmod, ripemd160, sha256, sha3}, 
	keywordstyle=[1]\color{blue}\bfseries,
	keywords=[2]{address, bool, byte, bytes, bytes1, bytes2, bytes3, bytes4, bytes5, bytes6, bytes7, bytes8, bytes9, bytes10, bytes11, bytes12, bytes13, bytes14, bytes15, bytes16, bytes17, bytes18, bytes19, bytes20, bytes21, bytes22, bytes23, bytes24, bytes25, bytes26, bytes27, bytes28, bytes29, bytes30, bytes31, bytes32, enum, int, int8, int16, int24, int32, int40, int48, int56, int64, int72, int80, int88, int96, int104, int112, int120, int128, int136, int144, int152, int160, int168, int176, int184, int192, int200, int208, int216, int224, int232, int240, int248, int256, mapping, string, uint, uint8, uint16, uint24, uint32, uint40, uint48, uint56, uint64, uint72, uint80, uint88, uint96, uint104, uint112, uint120, uint128, uint136, uint144, uint152, uint160, uint168, uint176, uint184, uint192, uint200, uint208, uint216, uint224, uint232, uint240, uint248, uint256, var, void, ether, finney, szabo, wei, days, hours, minutes, seconds, weeks, years},	
	keywordstyle=[2]\color{teal}\bfseries,
	keywords=[3]{block, blockhash, coinbase, difficulty, gaslimit, number, timestamp, msg, data, gas, sender, sig, value, now, tx, gasprice, origin},	
	keywordstyle=[3]\color{violet}\bfseries,
	identifierstyle=\color{black},
	sensitive=false,
	comment=[l]{//},
	morecomment=[s]{/*}{*/},
	commentstyle=\color{gray}\ttfamily,
	stringstyle=\color{red}\ttfamily,
	morestring=[b]',
	morestring=[b]"
}

\begin{document}

\title{Formally Verifying a Real World Smart Contract}

\author[1,2]{Alexandre Mota}
\author[2]{Fei Yang}
\author[2]{Cristiano Teixeira}
\affil[1]{acm@cin.ufpe.br\\Centro de Inform\'atica-UFPE, Av. Jornalista An\'{\i}bal Fernandes, s/n, Cidade Universit\'aria, Zip 50.740-560, Brazil}
\affil[2]{\{alexandre, fei, cristiano\}@lindylabs.net\\Lindy Labs,
https://www.lindylabs.net/, Portugal}

\maketitle

\begin{abstract}
 Nowadays, smart contracts have become increasingly popular and, as with software development in general, testing is the standard method for verifying their correctness. However, smart contracts require a higher level of certainty regarding correctness because they are difficult to modify once deployed and errors can result in significant financial losses. Therefore, formal verification is essential. In this article, we present our search for a tool capable of formally verifying a real-world smart contract written in a recent version of Solidity.
\end{abstract}


\section{Introduction}

 Nowadays, smart contracts have been attracting significant attention. While testing is the {\sl de facto} standard for verifying the correctness of smart contracts in software development, smart contracts require a more robust means of ensuring correctness because they are difficult to modify once deployed, and any flaws could result in substantial financial losses. Therefore, formal verification is necessary. This article showcases the endeavor to identify a tool capable of formally verifying a real-world smart contract written in Solidity.
 
 Until the end of March, 2022, Sandclock was verified mainly by Slither~\cite{slither19} and Echidna~\cite{echidna20}, besides independent auditors, a common practice nowadays in the Smart Contracts community. Slither is a static analysis solution that reports useful tips to improve a smart contract but it also reports several false alarms. And Echidna which performs fuzz testing and thus can take a long time to process sometimes; like any testing technique, Echidna is very useful when a bug is found because such a situation is certain but no so ever when no bug is found. In such a case, testers cannot affirm that the system is absent of bugs.
 
 Thus in April, 2022, we started a long journey to find out some tools (or at least a single one) that are able to formally verify Sandclock. We had a catalogue of expected properties already checked by Echidna. But a more trustworthy status was required.
 
 In September 2022, we finally arrived at a conclusion regarding the formal verification of Sandclock. This article details our journey, which included identifying limitations with certain tools and experiencing disappointment with others. We also discuss potential enhancements for some of the tools, and ultimately select the single tool capable of proving the desired properties for Sandclock.
 
 The main contributions of this article are:
 
 \begin{itemize}
     \item Investigating which testing and formal tools can deal with a real world smart contract written in Solidity 0.8.10;
     \item Discussing flaws in formal tools;
     \item Showing that Certora is the only formal tool able to formally verify Sandclock; our real-world system.
 \end{itemize}

The structure of this work is as follows: Section~\ref{sec:solidity} provides a brief introduction to the Solidity language, utilizing an excerpt of our real-world smart contract (Sandclock) as an example. Sandclock is then presented in more detail in Section~\ref{sec:Sandclock}. In Section~\ref{sec:investTools}, we discuss and present the various tools we attempted to use on Sandclock, finally finding the Certora prover as the best candidate. Section~\ref{sec:certora} introduces the Certora prover and its application to Sandclock in greater detail. Finally, in Section~\ref{sec:conclusion}, we summarize our conclusions and suggest potential future work.

\section{Solidity}
\label{sec:solidity}

Solidity~\cite{solidity22} is an object-oriented programming language designed for writing smart contracts executing in Ethereum Virtual Machine\footnote{https://ethereum.org/en/developers/docs/evm/}, which is widely supported by various blockchain platforms. The concept for Solidity was introduced by Ethereum co-founder Gavin Wood in 2014, and was further developed by Christian Reitwiessner, Ales Beregszaszi, and other Ethereum contributors. The initial release, version 0.1.2, was made available in August 2015, and since then, Solidity has been under continuous development with sponsorship from the Ethereum Foundation. The latest version of Solidity is 0.8.18, and a community of collaborators is involved in contributing to the language's evolution by adding new features, building systems, improving documentation, and addressing issues on GitHub. Solidity is influenced by popular programming languages such as C++, Python, and JavaScript, and is a statically typed, curly-braced, contract-oriented, high-level programming language. It includes common features such as inheritance, libraries for reusable code, and complex custom types. Listing~\ref{SandclockCode} displays an excerpt of Solidity code from our real world smart contract\footnote{The complete system can be found here: https://github.com/lindy-labs/sc\_solidity-contracts}.


\small
\begin{table}
\begin{lstlisting}[language=Solidity, label=SandclockCode, caption=Vault excerpt]
// SPDX-License-Identifier: UNLICENSED
pragma solidity >=0.8.10;

import {IERC20} from "@openzeppelin/contracts/token/ERC20/IERC20.sol";
...
import {CustomErrors} from "./interfaces/CustomErrors.sol";

contract Vault is IVault, ..., CustomErrors {
    using SafeERC20 for IERC20;
    ...
    using Counters for Counters.Counter;

    // Constants
    bytes32 public constant INVESTOR_ROLE = keccak256("INVESTOR_ROLE");
    ...
    uint256 public constant SHARES_MULTIPLIER = 1e18;

    // State
    /// @inheritdoc IVault
    IERC20Metadata public override(IVault) underlying;
    /// @inheritdoc IVault
    uint16 public override(IVault) investPct;
    ...
    /// The investment strategy
    IStrategy public strategy;
    /// Unique IDs to correlate donations that belong to the same foundation
    uint256 private _depositGroupIds;
    mapping(uint256 => address) public depositGroupIdOwner;
    ...

    constructor(IERC20Metadata _underlying, ..., SwapPoolParam[] memory _swapPools) {
        if (!_investPct.validPct()) revert VaultInvalidInvestpct();
        ...
        _grantRole(DEFAULT_ADMIN_ROLE, _admin);
        ...
        rebalanceMinimum = 10 * 10**underlying.decimals();
        _addPools(_swapPools);
        emit TreasuryUpdated(_treasury);
    }
\end{lstlisting}
\end{table}
\normalsize

At the beginning of each file, the ``pragma'' keyword is included to declare the Solidity version and enable compiler features. Within the ``contract'' section, similar to a class in OOP, attributes and methods are defined. Additionally, "event" is a member contract that can be inherited and functions as a method to be called later. Each time the "event" is emitted, it stores transaction log arguments that can be accessed with the contract's address [13]. Solidity also includes a special data type called "address" that represents a 20-byte Ethereum address. The "function" keyword is used to define executable blocks of code to modify the contract's state. As the smart contract is immutable, the ``constructor'' can only be called once during contract creation, and when it's finished executing, the code is deployed on the blockchain. As most of smart contracts manage crypto assets which have considerable monetary value, it's critical to verify its correctness. The security of smart contracts has been a major concern as the bugs in Solidity smart contracts have led to the hacking of assets worth billions of US dollars. These security issues in smart contracts highlight the significant role that developers play in creating vulnerabilities that can be exploited by attackers. Once the smart contract is deployed, it's impossible to modify the code, data, or logic. Hence, it's necessary for developers to perform testing on the code in testnet network before deploying it to the Mainnet, following best practices for all possible scenarios.

\section{Sandclock}
\label{sec:Sandclock}

\begin{figure}
    \includegraphics[width=\textwidth]{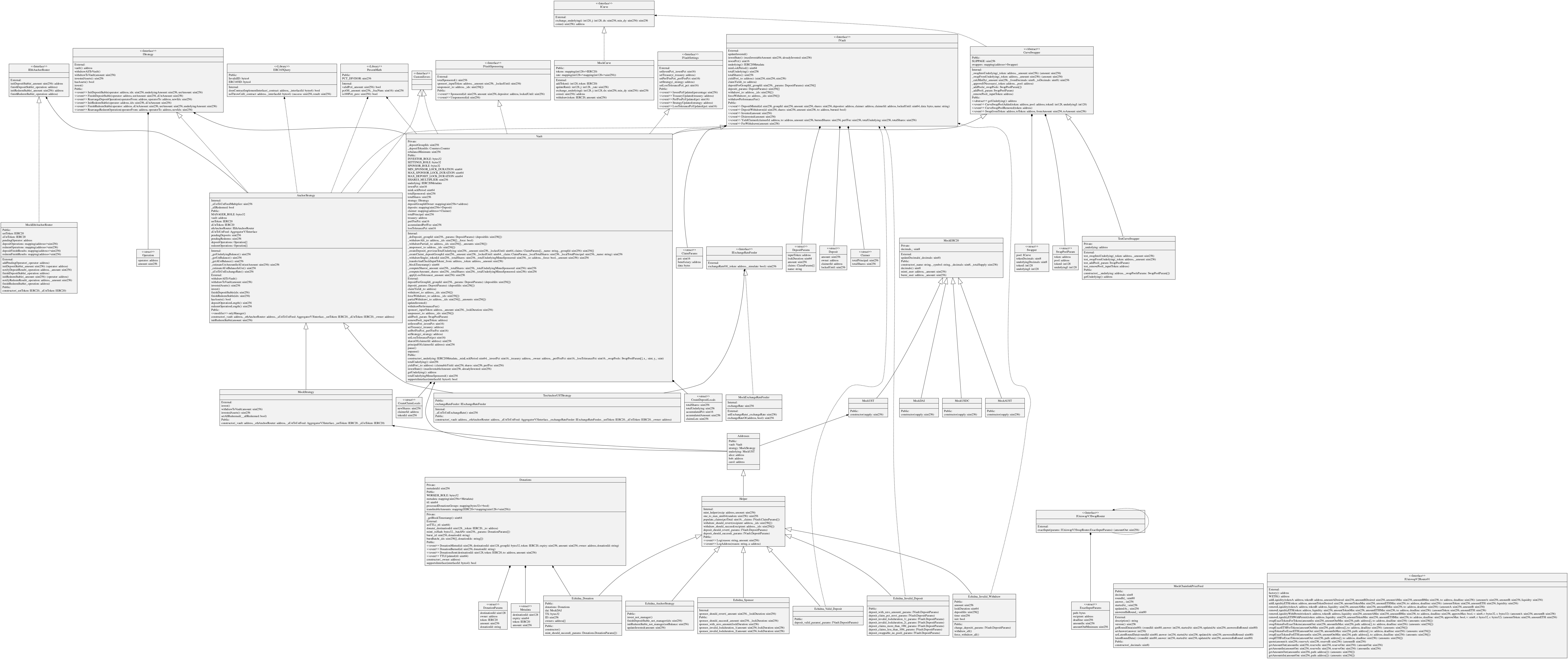}\hfill
    \caption{Sandclock (UML Class Diagram)}
    \label{fig:Sandclock}
\end{figure}

The entire Sandclock~\cite{Teixeira22} Solidity project (in the moment we are writing this article) can be seen in Figure~\ref{fig:Sandclock} as a UML class diagram. It has two core contracts: {\bf Vault} and {\bf Strategy}. The Vault is the contract users interact with to deposit, withdraw, claim yield, admin, etc., and the Strategy is the contract that invests the users’ funds to generate yield.

\subsection{Strategy}

Only a single strategy can be assigned to a vault, but any contract that implements the necessary interface can act as a strategy. It is possible to switch the strategy at any time by withdrawing all funds to the vault, updating the strategy's address, and rebalancing the vault.

All strategies used in the contracts interact with the same underlying asset. For example, if we deploy a strategy using Yearn's LUSD Vault, the strategy will use LUSD (Liquity USD) as the underlying asset, and the same applies to the vault. Nonetheless, the vault is integrated with Curve, which enables users to deposit various currencies. The underlying asset can be almost any other ERC20 that also functions as a stable coin.

Each strategy can be classified as either synchronous or asynchronous based on the immediacy of its interactions.

Synchronicity has more to do with deposits. For example, in an asynchronous strategy we have:

\begin{enumerate}
 \item Deposit;
 \item Share issuance;
 \item Investment (back-end invokes a function).
\end{enumerate}

The problem here is that the deposited funds are not ``at risk'' upon executing functions of steps 1. and 2. But they receive yield! They are only ``at risk'' in functions of step 3. In a synchronous strategy, all steps occur atomically.

All strategies have instantaneous withdrawals, through a reserve that we keep ``uninvested''. Otherwise, they would also be asynchronous.

\subsection{Vault}

The vault is the centerpiece of the system. It

\begin{itemize}
\item keeps track of deposits, sponsors, and yield;
\item decides how much to invest or disinvest in the strategy;
\item applies the performance fee to the yield;
\item handles most administrative tasks.
\end{itemize}

Regular users of Sandclock will interact with the vault by depositing, withdrawing, and claiming yield. While the vault's strategy uses the same underlying asset, users can deposit in other currencies. In addition to depositing and withdrawing funds, depositors can assign the yield generated to anyone, including themselves, another account, or Sandclock's treasury in case of a donation.

When creating a vault, the investment percentage parameter determines how much of the total funds will be invested through the strategy. The remaining funds are held as reserves by the vault to support instantaneous withdrawals and yield claims.

The vault offers support for both synchronous and asynchronous strategies. In synchronous mode, withdrawals are promptly processed provided that the initial locked period on the vault has elapsed. If a user wishes to withdraw an amount that surpasses the reserves held by the vault, the vault will promptly rebalance with the selected strategy, and the withdrawal will proceed accordingly.

When operating in asynchronous mode, any attempts to withdraw or claim funds will be unsuccessful if the vault's reserves do not possess sufficient funds. Under these circumstances, users are required to await the back-end to rebalance the vault with the designated strategy, thereby making additional funds available.

\section{Investigated tools}
\label{sec:investTools}

Our mining and testing process was based mainly on~\cite{repos22} and Google Scholar.

\begin{itemize}
    \item eThor~\cite{ethor20}: This tool needs a specific version of Z3~\cite{z308} to work and it took several hours to check Sandclock without any answer. We contacted the authors, which were very kind, but we gave up due to such difficulties;
    \item Vandal~\cite{vandal18}: It found ``issues'' on Sandclock. But even contacting the authors we cannot trace back the findings to the source code. Besides that, Mythril did not find any issue and Vandal's authors said that Mythril is more precise concerning reentrancy issues;
    \item Maian~\cite{maian22}: It analyzed the contract easily using a docker installation. No problem was reported;
    \item SmartBugs~\cite{smartbugs20} allowed us to quickly test HoneyBadger~\cite{honeybadger19}, Oyente~\cite{oyente16}, Manticore~\cite{manticore19}. During the verification of Sandclock, Manticore appeared to experience a freeze (subsequently, after waiting for several hours, we decided to abandon the verification process), while the other tools reported that everything was functioning properly. Curiously, SmartBugs cannot run Mythril~\cite{mythril22} but we can run it directly;
    \item SKLEE~\cite{sklee22}: Interesting and recent solution to check Solidity via a translation to C and reuse of KLEE. But it also has issues concerning Solidity version and its implementation has several limitations;
    \item We tested Securify~\cite{securify18}, Solidifier~\cite{solidifier21}, solc-verify~\cite{solcverify19}, and SmartACE~\cite{smartace22}. But they all needed Solidity $<$0.5 (and some Solidity $<$0.6);
    \item SMTChecker~\cite{smtchecker18}: This was the initial formal tool used for Sandclock, but the first attempt was disappointing as it relied on all properties and the Solidity compiler used by Sandclock (based on hardhat\footnote{https://hardhat.org/}). To address this, we contacted the authors and utilized a static version of {\tt solc} for Mac, focusing only on assertions. However, we encountered different behaviors when comparing the original project (several folders and files) and its flat version, ultimately finding the flat version to be more reliable. Despite significant effort, we discovered that SMTChecker struggled with {\tt keccak256} constants (see Listing~\ref{SandclockCode}, line~14) and misbehaving just at the constructor of Vault, even reporting the obvious assertion {\tt assert(true)} as non-satisfiable. As SMTChecker was the only available tool for verifying Sandclock, we drew on prior knowledge and expertise~\cite{Betal04,DMS11} to simplify, prune, and refactor the code as much as possible, reducing it from the original 3997 LOC to a 1663 LOC version after several hours of manual effort. Despite significant manual effort and hand inlining of code, the first property of Sandclock was eventually proven. We also attempted to use HEVM~\cite{hevm22} for further transformations, but encountered difficulties with EVM bytecode deviations even with minor syntactic changes;
    \item Verismart~\cite{verismart20}: We tested Verismart almost simultaneously with the formal verifier SMTChecker. Just like with SMTChecker, we needed a flattened version of Sandclock and had to eliminate all events, such as the contract {\tt CustomErrors} (refer to Listing~\ref{SandclockCode}, lines~6 and~8). However, unlike SMTChecker, we did not need to simplify, prune or refactor Sandclock. Verismart was able to handle it as is. Our initial approach was to verify multiple generic properties at once, ranging from arithmetic to ether leak. The outcome was
    \begin{lstlisting}[language=Solidity, caption=Vault excerpt]
============ Statistics ============
# Iter                    : 1
# Alarm / Query           : 46 / 66
- integer over/underflow  : 42 / 60
- division-by-zero        : 3 / 5
- kill-anyone             : 0 / 0
- ether-leaking           : 1 / 1

Time Elapsed (Real) : 4436.64151192
Time Elapsed (CPU)  : 4288.250747
\end{lstlisting}

Although this report is interesting, it also reveals certain weaknesses. For example (where {\tt DZ} means possible division by zero),
\small
\begin{verbatim}
...
[62] [DZ] line 1666, ((_amount * _fracNum) / PCT_DIVISOR) : unproven
...
\end{verbatim}
\normalsize

But the constant {\tt PCT\_DIVISOR} is defined inside {\tt PercentMath} as follows. This is the kind of manual inlining we had to perform sometimes to get a more trustworthy result.

\begin{lstlisting}[language=Solidity, firstnumber=1644, caption=Vault excerpt]
library PercentMath {
    // Divisor used for representing percentages
    uint256 public constant PCT_DIVISOR = 10000;
    ...
}
\end{lstlisting}

A similar issue arose with SMTChecker as well. Our experience with Verismart was highly disappointing and ultimately led us to abandon it. This occurred when we attempted to verify whether the variable "value" could have different values at distinct points in the Solidity code without any assignments between these locations. The report we received from Verismart was the reason for our decision to move on from it.

\begin{lstlisting}[language=Solidity, firstnumber=1, caption=Report]
[1] [ASSERT] line 86, (value == 1) : proven
[2] [ASSERT] line 93, (balance_this_after == (balance_this_before + value)) : proven
[3] [ASSERT] line 94, (balance_this_after == 0) : unproven
[4] [ASSERT] line 95, (value == 0) : proven
\end{lstlisting}

    \item Halmos\footnote{https://github.com/a16z/halmos}: This is a very recent Symbolic Bounded Model Checker for Ethereum Smart Contracts Bytecode.
    \begin{itemize}
        \item Symbolic: By using symbolic function arguments and storage states, Halmos can execute the provided contract bytecode and systematically investigate all feasible behaviors of the contract;
        \item Bounded: Halmos has the ability to automatically unroll loops up to a predetermined limit and define the size of arrays with variable length, making it possible to run without any additional user annotations;
        \item Model Checking: Halmos can either demonstrate that assertions are never infringed by any inputs or produce a counter-example, which makes it suitable not only for formal verification of the contract but also for bug detection.
    \end{itemize}
    Regrettably, Halmos is still in development, and during my initial usage, it encountered an error related to ``ValueError: invalid literal for int() with base 10: "sha3\_var1"''. The issue is currently being resolved by the developers;

    \item The Certora Prover~\cite{certora22}: This was the sole tool able to formally verify Sandlock with trustworthiness and minimum effort. Section~\ref{sec:certora} presents its main advantages and why we think it worked so well with Sandclock;
    \item We thought to use Klab\footnote{https://github.com/dapphub/klab} almost concurrently with Certora. But after examining this repository\footnote{https://github.com/dapphub/k-dss/}, we saw that KLab was used on a manually created version of the Multi-Collateral DAI\footnote{https://github.com/makerdao/dss}, abstracting a lot of part of the real-world system, and with its last update 4 years ago. Thus, with a little further investigation we saw that MakerDAO, which manages the stable coin Multi-Collateral DAI, replaced KLab for the Certora prover\footnote{https://hackmd.io/@SaferMaker/DAICertoraSurprise}, and this reinforced our conclusions that the Certora prover is the right tool to this kind of job.
\end{itemize}

\section{Verifying Sandclock with the Certora prover}
\label{sec:certora}

The Certora prover~\cite{certora22}, like solc-verify~\cite{solcverify19} and Solidifier~\cite{solidifier21}, uses a specification language (CVL---Certora Verification Language, in the case of Certora) apart from the programming language (Solidity). Certora appears to have designed CVL to share many similarities with Solidity, likely as a tactic to encourage developers to use their prover with greater ease and skill.

\begin{figure}
    \includegraphics[width=\textwidth]{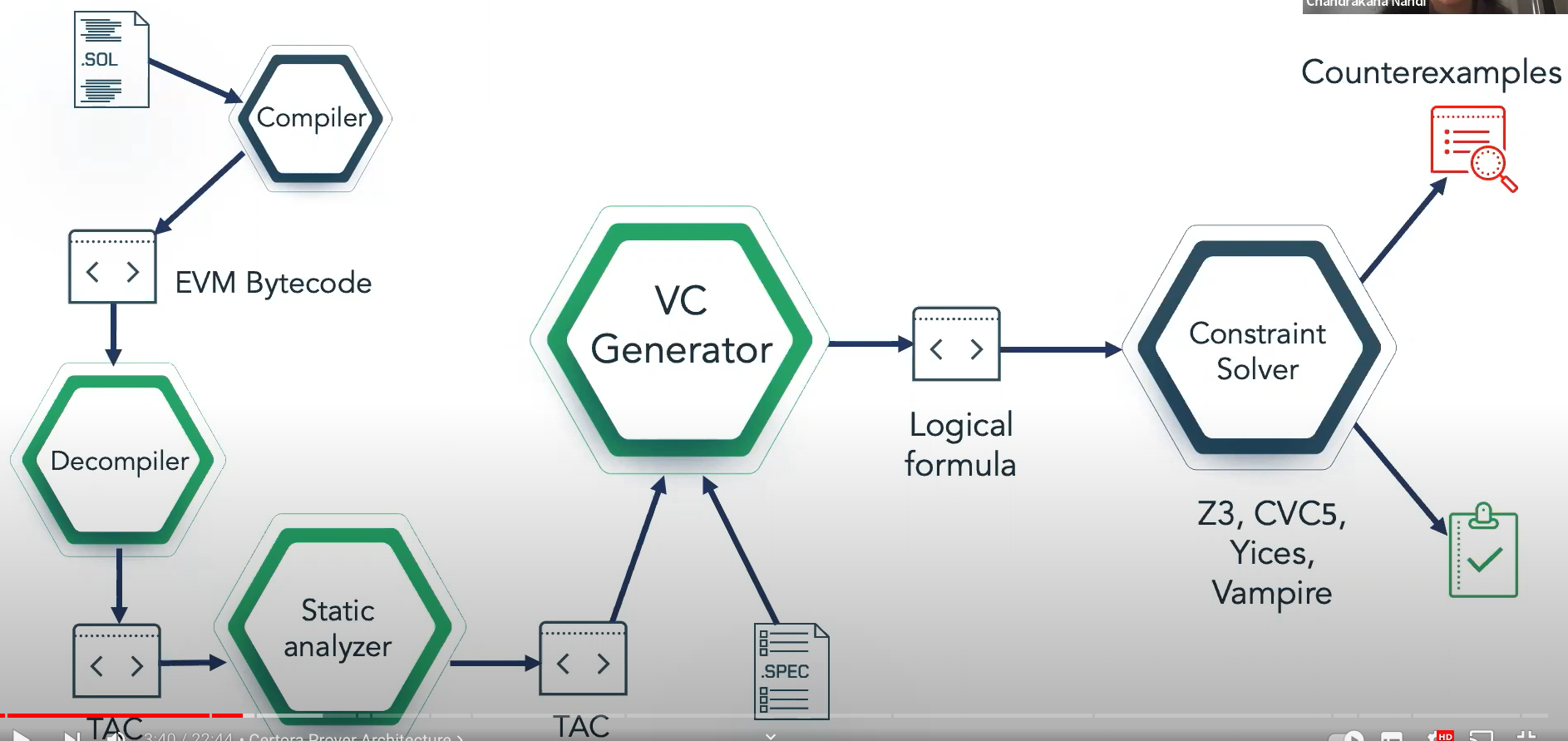}\hfill
    \caption{The Certora Prover Architecture}
    \label{fig:TheCertoraProver}
\end{figure}

The Certora prover provides a more advanced approach to Smart Contract verification. It automates many of the manual processes required when using SMTChecker and Verismart. As we can see in Figure~\ref{fig:TheCertoraProver}.

The Certora client---a python program---receives 2 main inputs from the user:

\begin{itemize}
 \item the Solidity smart contract to verify;
 \item The specification of the properties of the contract.
\end{itemize}

The specification for verification is written in CVL (Certora Verification Language), and the client allows for numerous parameters to customize the behavior of the verifier. Due to the potentially extensive nature of the command, it is recommended to save the command as a shell script and execute the script to invoke it.

The Certora prover simplifies the process of verifying a smart contract in a professional manner. Firstly, the prover compiles the Solidity smart contract to obtain EVM bytecodes. These bytecodes are then decompiled into a simplified intermediary three address code (TAC) language using static analysis, which removes irrelevant variables for the property being analyzed. The specification is used to generate verification conditions for the simplified code, which are translated into logical formulas in the SMT-LIB format. These formulas are fed into various solvers, each with different capabilities, to find values for variables. The negation of the intended property is checked to determine if a solution is found, which represents a counterexample, or if the property is proven. The use of bytecode and the simplification performed by the static analyzer are the two main advantages of the Certora prover over other verification tools, making it more scalable and capable of analyzing real-world smart contracts. Figure~\ref{fig:TheCertoraProver} illustrates the architecture of the Certora prover.

\subsection{How The Certora Prover is supposed to be used}

The final paragraph in the previous section is designed to be transparent to the end user, enabling them to concentrate on the properties of interest rather than the limitations of SMT solvers in dealing with the intricacies of a programming language and a real-world system. The end user is trained to use the Certora prover as follows:

\begin{itemize}
    \item Have a good understanding of the Solidity smart contract to be analyzed;
    \item List all state variables of the core contracts to be analyzed;
    \item List all static global settings (they are also seen as state variables);
    \item List all external/public functions that change state variables of the core contracts;
    \item List all external/public functions that are privileged and change settings;
    \item List all external/public functions that are privileged and move underlying assets;
    \item List all external/public functions that are view only and change nothing.
\end{itemize}

The next step is to create a table similar to the one below.

\begin{table}[h]
\begin{tabular}{|l|l|l|l|l|l|l|}
\hline
{\bf N$^o$} & {\bf Property} & {\bf Cat.} & {\bf Prio.} & {\bf Spec.} & {\bf Ver.} & {\bf Link} \\ \hline
 \vdots & \vdots  & \vdots & \vdots & \vdots & \vdots & \vdots \\ \hline
 8 & \parbox{4cm}{`deposit(...)` the underlying token should reduce the user's balance by the specified amount while increasing Vault's balance by the same amount. It should also increase claimer's `totalShares` and the Vault's `totalShares` by the same amount.} & var. trans. & high & Y & Y & https:\ldots \\ \hline
 \vdots & \vdots  & \vdots & \vdots & \vdots & \vdots & \vdots \\ \hline
\end{tabular}
\end{table}

Once the table mentioned above is created, the subsequent step is to convert it into the CVL language. To overcome any limitations of CVL, such as complex data types, Solidity programming can be used to create a wrapper for the original code. All Certora encoding, including the harness and CVL specifications, can be stored in a certora folder at the same level as the smart contract, along with the scripts.

\subsection{Using The Certora Prover on Sandclock}

After gaining some understanding of the Sandclock smart contract, we focus on verifying the Vault smart contract, which serves as the core contract. We compile a list of all the relevant elements based on the previous explanation. The Vault includes:

\begin{enumerate}
    \item State variables:
    \begin{itemize}
        \item {\tt totalSponsored} (type {\tt uint256}): total sponsored amount of the underlying ERC20 asset;
        \item {\tt totalShares} (type {\tt uint256}): total shares of the all the users;
        \item \ldots
    \end{itemize}
    \item Static global settings (also state variables):
    \begin{itemize}
        \item {\tt lossTolerancePct} (type {\tt uint16}): loss tolerance percentage. It's specified in the constructor but can be updated by a settings account;
        \item {\tt investPct} (type {\tt uint16}): the percentage to invest by the Vault. Vault may leave some users' funds uninvested as reserve for user to withdraw anytime. It's specified in the constructor but can be updated by a settings account;
        \item \ldots
    \end{itemize}
    \item External/public functions that change state variables:
    \begin{itemize}
        \item {\tt function deposit(DepositParams calldata\_params) external\\ nonReentrant whenNotPaused returns (uint256[] memory depositIds)};
        \item {\tt function depositForGroupId(uint256 \_groupId, \\DepositParams calldata\_params) external nonReentrant \\whenNotPaused returns (uint256[] memory depositIds)};
        \item \ldots
    \end{itemize}
    \item External/public functions that are privileged and change settings:
    \begin{itemize}
        \item {\tt function transferAdminRights(address \_newAdmin) external \\onlyAdmin};
        \item {\tt function pause() external onlyAdmin};
        \item \ldots
    \end{itemize}
    \item External/public functions that are privileged and move underlying assets:
    \begin{itemize}
        \item {\tt function updateInvested() external override(IVault) onlyKeeper};
        \item {\tt function withdrawPerformanceFee() external override(IVault) \\onlyKeeper}.
    \end{itemize}
    \item External/public functions that are view only and change nothing:
    \begin{itemize}
        \item {\tt function investState() public view override(IVault) returns \\(uint256 maxInvestableAmount, uint256 alreadyInvested)};
        \item {\tt function paused() public view virtual returns (bool)} (inherited from Pausable.sol).
    \end{itemize}
\end{enumerate}

\begin{figure}[h]
    \includegraphics[width=\textwidth]{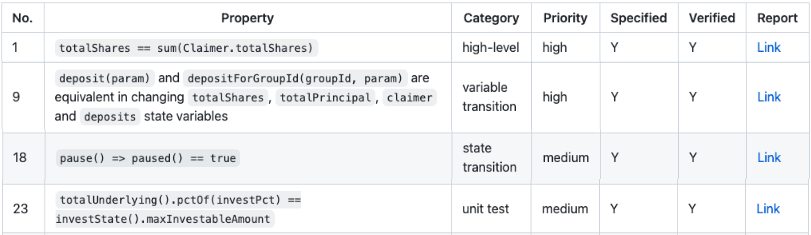}\hfill
    \caption{Sandclock Properties excerpt}
    \label{fig:SandclockProperties}
\end{figure}

Now we have to enumerate expected properties from the Vault smart contract. An excerpt of all properties (the total is 34) can be seen in Figure~\ref{fig:SandclockProperties}.

The final step involves populating the harness and spec folders and running the Certora prover. In the case of Sandclock, the creation of the `VaultHarness.sol' file is crucial in the harness folder. This file inherits from `Vault.sol' and includes adaptation code to establish a link between the CVL spec and the Vault smart contract. For instance:

\tiny
\begin{table}[h]
\label{lst:Sandclock}
\begin{lstlisting}[language=Solidity, caption=VaultHarness excerpt]
...
contract VaultHarness is Vault {
...
    function deposit(address inputToken, uint64 lockDuration, 
        uint256 amount, uint16[] calldata pcts,
        address[] calldata beneficiaries, bytes[] calldata datas,
        uint256 slippage
    ) external nonReentrant whenNotPaused {
        DepositParams memory params = buildDepositParams(
            inputToken, lockDuration, amount, pcts, beneficiaries,
            datas, slippage);
        this.deposit(params);
    }
...
    function buildDepositParams(address inputToken, uint64 lockDuration, 
        uint256 amount, uint16[] calldata pcts,
        address[] calldata beneficiaries, bytes[] calldata datas,
        uint256 slippage
    ) internal returns (DepositParams memory) {
        require(pcts.length == beneficiaries.length && pcts.length == datas.length);
        ClaimParams[] memory claims;
        for(uint256 i = 0; i < pcts.length; i++) {
            claims[i] = ClaimParams({pct: pcts[i], beneficiary: beneficiaries[i],
                data: datas[i]
            });
        }
        return DepositParams({inputToken: inputToken, lockDuration: lockDuration,
            amount: amount, claims: claims, name: "test", slippage: slippage
        });
    }
...
}
\end{lstlisting}
\end{table}
\normalsize

Note that the function {\tt deposit} in {\tt VaultHarness} provides a CVL supported way to deal with the complex data structure {\tt DepositParams}, which is created through the function {\tt buildDepositParams}.

And the final step before checking a property using the Certora prover is obviously to have such a property in CVL terms. Thus, suppose Property~1 from the table in Figure~\ref{fig:SandclockProperties}. It can be stated in CVL as.

\scriptsize
\begin{verbatim}
invariant tatalShares_equals_sum_of_claimer_shares()
    totalShares() == sumOfClaimerShares()
\end{verbatim}
\normalsize

It is worth noting that the function {\tt totalShares} is already available, while {\tt sumOfClaimerShares} is not. To calculate {\tt sumOfClaimerShares}, a {\tt for} loop needs to be performed on the elements of a structure. However, SMT solvers do not handle loops easily, so CVL offers a workaround. Since CVL operates at the EVM bytecode level, it provides {\tt ghost} functions and {\tt hook}s. The {\tt ghost} functions are used to store values, while the {\tt hook}s are used to intercept the EVM bytecode when load or store instructions are being processed by the EVM.

Then the function {\tt sumOfClaimerShares} becomes a ghost function as follows.

\scriptsize
\begin{verbatim}
ghost sumOfClaimerShares() returns uint256   {
    init_state axiom sumOfClaimerShares() == 0;
}
\end{verbatim}
\normalsize

The function's body indicates that its value is initially set to {\tt 0}, which is significant because the default initial value for a {\tt uint256} variable in Solidity is zero. The {\tt hook} entry, however, is a bit more complicated. It is based on an old tradition in Formal Methods that uses the current state of a variable (suffix {\tt @new}) and its previous state (suffix {\tt @old}) to establish the relationship between them. Here is the statement:

\scriptsize
\begin{verbatim}
hook Sstore claimer[KEY address k].(offset 32) uint256 amount (uint256 oldAmount) STORAGE {
    havoc sumOfClaimerShares assuming
        sumOfClaimerShares@new() == sumOfClaimerShares@old() + (amount - oldAmount);
}
\end{verbatim}
\normalsize

The above operation focuses on the state variable {\tt claimer}

\footnotesize    
\begin{verbatim}
mapping(address => Claimer) public claimer
\end{verbatim}
\normalsize

\noindent Particularly in its field element {\tt totalShares} of its related structure.
 
\scriptsize
\begin{verbatim}
struct Claimer {
    uint256 totalPrincipal;
    uint256 totalShares;
}
\end{verbatim}
\normalsize

The last step of the Formal Verification campaign is to inform the findings (List of Issues Discovered) obtained by running Certora on the specified rules:

Severity: Low

\begin{tabular}{|l|l|}
\hline
   Property violated & \parbox{11cm}{When user withdraws principal, the requested amount should be sent to the user} \\ \hline
   Severity & Low \\ \hline
   Issue  & Dust in the Vault \\  \hline
   Description  & \parbox{11cm}{When user withdraw the principal from the vault, the vault may leave some dust due to rounding of uint division in Solidity. E.g., user withdraws 1000 LUSD, the vault may send only 999.99999... LUSD if the share price is not exactly an integer, which can happen when underlying asset value has changed, e.g., yield earned. A counter example can be found in https://prover.certora.com/output/52311/0ca7e4fb0\\86d308eeb51?anonymousKey=2a093f38c9406a97d4b\\9d56e746f07aec63616a8} \\ \hline
   Mitigation/Fix & \parbox{11cm}{update the `\_withdrawSingle` internal function to remove the redundant and imprecise `computeAmount` function call and return the `\_amount` value directly} \\
\hline
\end{tabular}

\section{Conclusions}
\label{sec:conclusion}

As the popularity of smart contracts continues to grow, ensuring their correctness has become increasingly important. While testing is a common approach, it may not be sufficient due to the immutability of deployed smart contracts and the potential for significant financial losses if flaws are present. Formal verification is therefore necessary.

In this article, we shared our journey to identify a tool capable of formally verifying a real-world smart contract written in Solidity (specifically, version 0.8.10). After experiencing limitations and disappointment with various tools, we ultimately found that Certora was the only tool capable of proving the desired properties for Sandclock.

To the best of our knowledge, no previous work has attempted to formally verify a really complex real-world smart contract as we have done in this study. However, we did discover several related works that focused on the tools discussed and analyzed in Section~\ref{sec:investTools}.

In our upcoming work, we plan to apply Certora to additional strategies developed by Lindy Labs for Sandclock. Furthermore, we aim to train engineers to utilize Certora effectively in the future. In addition, we plan to integrate Certora with testing tools like Echidna and Foundry. While fuzz testing can be impractical in certain scenarios, Certora can provide better performance. However, Certora may time out when verifying specific properties, making it necessary to supplement it with other testing approaches.

Apart from Solidity, Lindy Labs also uses Cairo\footnote{https://starkware.co/cairo/} as the programming language to create another real world smart system named Aura\footnote{https://github.com/lindy-labs/aura\_contracts}. Another future work is creating a formal verification machinery able to verify Aura.

\paragraph{{\bf Acknowledgements}} We would like to thank all people from Lindy Labs and, in particular, Gabriel Po\c{c}a (for clarifying certain things about Sandclock) and Miguel Palhas (for allowing us to use Certora initially). Alexandre Mota would like to thank CNPq for grant Number 307824/2021-7.

\bibliographystyle{elsarticle-num}

\bibliography{article}

\end{document}